\documentclass[letterpaper]{JHEP3}
\usepackage{amssymb,amsfonts}
\usepackage{cite}
\usepackage{graphicx}
\usepackage{latexsym}

\newcommand{\CF}{{\cal F}}

\newcommand{\CO}{{\cal O}}

\newcommand{\bx}{{\bf x}}

\newcommand{\p}{\partial}

\newcommand{\diff}{{\rm Diff}}

\renewcommand{\tilde}[1]{\widetilde{#1}}
\newcommand{\be}{\begin{equation}}
\newcommand{\ee}{\end{equation}}
\newcommand{\bea}{\begin{eqnarray}}
\newcommand{\eea}{\end{eqnarray}}
\newcommand{\eg}{{\it e.g.}}
\title{Anisotropic Conformal Infinity}
\author{Petr Ho\v{r}ava and Charles M. 
Melby-$\!$Thompson\\
Berkeley Center for Theoretical Physics and Department of Physics\\
University of California, Berkeley, CA, 94720-7300\\
and\\
Theoretical Physics Group, Lawrence Berkeley National Laboratory\\
Berkeley, CA 94720-8162, USA}
\abstract{We generalize Penrose's notion of conformal infinity of spacetime, 
to situations with anisotropic scaling. This is relevant not only for 
Lifshitz-type anisotropic gravity models, but also in standard general 
relativity and string theory, for spacetimes exhibiting a natural 
{\it asymptotic\/} anisotropy.  Examples include the Lifshitz and 
Schr\"odinger spaces (proposed as AdS/CFT duals of nonrelativistic field 
theories), warped $AdS_3$, and the near-horizon extreme Kerr geometry.  
The anisotropic conformal boundary appears crucial for resolving 
puzzles of holographic renormalization in such spacetimes.}  
\begin{document}
%%%%%%%%%%%%%%%%%%%%%%%%%%%%%%%%%%%%%%%%%%%%%%%%%%%%%%%%%%%%%%%%%%%%%%%%%%%%%%%
\section{Introduction}

Recently, string theory and quantum gravity have begun to expand into 
territories traditionally associated with theoretical condensed matter 
physics.  In the process, the apparent divide between relativistic and 
nonrelativistic systems is becoming significantly blurred.  For example, 
relativistic gravity solutions have been proposed as duals of nonrelativistic 
quantum field theories (NRQFTs) \cite{son,mcg} characterized by anisotropic 
scaling of time and space,
\be
\label{scaling}
t\to\lambda^zt,\qquad x^i\to\lambda x^i,
\ee
with dynamical exponent $z\neq 1$.  In another development, gravity 
models have been proposed \cite{mqc,lif,spdim} in which the gravitational 
field itself is subject to anisotropic scaling (\ref{scaling}) at short 
spacetime distances, leading to an improved ultraviolet behavior.  

At this new interface of condensed matter with quantum gravity, 
challenges and puzzles emerge.  For example, extending the concept of 
holographic renormalization (see, \eg, \cite{holo} for a review) to 
nonrelativistic QFTs has proven surprisingly difficult.  In standard 
holographic renormalization, the counterterms in a relativistic field theory 
are constructed from the analysis of the asymptotic behavior of bulk gravity 
near the boundary of spacetime.  Many of the difficulties with holographic 
renormalization of NRQFTs can be traced to the fact that the proposed gravity 
duals have a degenerate conformal boundary, as defined in the sense of Penrose 
\cite{gkp,penrin}.  This degenerate behavior indicates that Penrose's 
definition of conformal infinity is insufficient to handle holography in such 
spacetimes, and that it needs to be generalized to incorporate systems with 
anisotropic scaling.  

In this paper, we present such a generalization of conformal 
infinity of spacetime.  Our construction is based on concepts first developed 
in the context of quantum gravity with anisotropic scaling \cite{mqc,lif}.  
Here we focus on the main idea of the construction, illustrated by a few 
examples; further details will appear elsewhere \cite{sch}.

\section{Anisotropic Conformal Infinity: The Spatially Isotropic Case}

One feature common to geometries dual to NRQFTs is that their asymptotic 
behavior ``near the boundary'' reflects the anisotropic scaling 
(\ref{scaling}) of the dual NRQFT\@.  This suggests that the correct 
notion of asymptopia and conformal infinity should reflect this anisotropy 
in the conformal transformations near the boundary.  However, the idea of 
using anisotropic conformal transformations to define the boundary of 
spacetime immediately leads to apparent conflicts:  The conformal boundary 
must be a geometric object, defined such that it is preserved by the 
symmetries of gravity; but spacetime diffeomorphisms only allow isotropic 
Weyl transformations, reducing us to Penrose's original definition.  

\subsection{The main idea}

The observation crucial for resolving these conflicts was made \cite{mqc} 
in gravity models with anisotropic scaling:  Appropriately defined local 
anisotropic Weyl transformations are compatible with the restricted group 
$\diff(M;\CF)$ of those diffeomorphisms of spacetime $M$ that preserve a 
preferred foliation $\CF$ of $M$ by fixed time slices.  This fact allows us to 
define the concept of {\it anisotropic conformal infinity}, which legitimizes 
the asymptopia of many spacetimes, including those that appeared as duals of 
NRQFTs.  

In the anisotropic gravity models of \cite{mqc,lif}, the reduction of 
symmetries to $\diff(M;\CF)$ is a consequence of the gauge symmetries of the 
system.  However, our construction of anisotropic conformal infinity is valid 
beyond the context of \cite{mqc,lif}, and applies naturally to a large class 
of solutions of standard general relativity and string theory:  It is 
sufficient that the symmetries reduce to $\diff(M;\CF)$ {\it only 
asymptotically, near the spacetime boundary}.  As we will see, this is indeed 
the behavior exhibited by the gravity duals of NRQFTs.  This shows that the 
ideas of \cite{mqc,lif} find meaningful applications beyond the context of 
anisotropic gravity models.

The group $\diff(M;\CF)$ of foliation-preserving diffeomorphisms is generated 
by
\be
\xi\equiv f(t)\p_t+\xi^i(t,x^j)\p_i.
\ee
$\diff(M;\CF)$ appeared in \cite{mqc,lif} as the gauge symmetry of gravity 
with anisotropic scaling (\ref{scaling}).  In the canonical (ADM) 
parametrization of the metric,
\be
ds^2=-N^2dt^2+g_{ij}(dx^i+N^idt)(dx^j+N^jdt),
\ee
the $\diff(M;\CF)$ generators act via
\bea
\delta_\xi N&=&f\dot N+\dot f N+\xi^i\p_iN,\cr
\delta_\xi N_i&=&f\dot N_i+\dot f N_i+\xi^j\p_jN_i+\p_i\xi^jN_j+\dot\xi^j
g_{ij},\cr
\delta_\xi g_{ij}&=&f\dot g_{ij}+\xi^k\p_kg_{ij}+\p_i\xi^kg_{jk}
+\p_j\xi^kg_{ik}.
\eea
Using an arbitrary smooth nonzero scale factor $\Omega(t,x^i)$, we define 
the anisotropic Weyl transformations to be
\be
\label{wl}
\tilde N=\Omega^{z}N,\qquad \tilde g_{ij}=\Omega^2 g_{ij},\qquad \tilde N_i
=\Omega^2 N_i.
\ee
It was observed in \cite{mqc,lif} that the generators $\delta_\omega$ of these 
anisotropic Weyl transformations form a closed algebra with the generators of 
$\diff(M;\CF)$:
\be
[\delta_\xi,\delta_\omega]=\delta_\varpi,\qquad{\rm with}\quad
\varpi=f\dot\omega+\xi^i\p_i\omega.
\ee
Given (\ref{wl}), our definition of anisotropic conformal infinity of 
spacetime $M$ with metric $ds^2$ is essentially the same as in the 
isotropic case:  We map $M$ by an anisotropic Weyl transformation $\Omega$ to 
an auxiliary spacetime $\tilde M$ with a rescaled metric $\tilde{ds^2}$, 
choosing $\Omega$ such that the region near infinity in $M$ is mapped to 
points inside a compact region of $\tilde M$.  Under this map, the {\it ideal 
points at anisotropic conformal infinity of $M$} correspond to the boundary of 
the image of $M$ inside $\tilde M$, where the scale factor $\Omega$ vanishes 
while $d\Omega\neq 0$.  We will denote the anisotropic conformal infinity of 
$M$ by $\p M$.

\subsection{Asymptotic structure of the Lifshitz space}

Our first example is the Lifshitz spacetime, with metric
\be
\label{lifmet}
ds^2=-\frac{dt^2}{w^{2z}}+\frac{d\bx^2+dw^2}{w^2}.
\ee
This geometry was proposed in \cite{klm} as the gravity dual for NRQFTs with 
Lifshitz-type scaling without Galilean invariance.  With the choice of 
$\Omega=w$, we find that the Lifshitz spacetime is anisotropically conformal 
to the portion of the flat spacetime with $w>0$, with the standard metric
\be
\label{lifmets}
\tilde{ds^2}=-dt^2+d\bx^2+dw^2.
\ee
The anisotropic conformal boundary at infinity in (\ref{lifmet}) is mapped to 
$w=0$ in (\ref{lifmets}).  In this example, we see that

($a$) the anisotropic boundary is of codimension one, 

($b$) the bulk metric induces an anisotropic conformal class of metrics in the 
boundary, and 

($c$) the action of the anisotropic conformal symmetry in the boundary is 
induced from the action of the bulk isometries.  

Point ($c$) deserves a closer explanation:  In analogy with the isotropic 
case, we define {\it anisotropic conformal transformations} of a fixed metric 
on $\p M$ to be those $\diff(\p M;\CF)$ transformations that map the metric to 
itself up to an anisotropic Weyl transformation.  Here $-dt^2+d\bx^2$ is a 
representative of the anisotropic conformal class of metrics at the boundary 
of the Lifshitz space.  The corresponding group of anisotropic conformal 
transformations is finite-dimensional, generated by time translations, 
spatial translations and rotations, and the anisotropic scaling transformation 
(\ref{scaling}).  It is this conformal symmetry group whose action on $\p M$ 
is induced from the bulk isometries of the Lifshitz space $M$.

\section{Spatially Anisotropic Conformal Infinity}

Our other examples require a more refined structure, with several dynamical 
exponents and with nested foliations of spacetime.  We consider the case with 
scaling
\be
\label{scaledbl}
t\to\lambda^zt,\qquad x^i\to\lambda x^i,\qquad y^a\to\lambda^{\zeta}y^a,
\ee
and look for anisotropic Weyl transformations which reduce for a constant 
$\Omega\equiv\lambda$ to (\ref{scaledbl}), and form a closed group with 
those spacetime diffeomorphisms $\diff(M;\CF_2)$ that preserve the structure 
of a nested foliation $\CF_2$ of spacetime.  $\diff(M;\CF_2)$ is generated by 
\be
\label{foldbl}
\xi\equiv f(t)\p_t+\xi^i(t,x^j)\p_i+\eta^a(t,x^j,y^b)\p_a.
\ee
One could first assume that $g_{ab}$ is invertible, and simply iterate the 
logic from the single-foliation case.  Examples with this behavior would 
include the obvious generalizations of the Lifshitz spacetime, with 
an additional spatial anisotropy and scaling (\ref{scaledbl}); the 
anisotropic conformal infinity of such spacetimes again exhibits the same 
features ($a$)-($c$) as in the single-foliation Lifshitz space. 

We will be interested in a different class of examples, in which $g_{ab}$ 
is not necessarily invertible.  The most interesting case corresponds to 
$\zeta=0$; spatial dimensions with this scaling will be called 
``ultralocal.''  We specialize to the case of just one ultralocal 
dimension $y$, and parametrize the metric as
\bea
\label{parmet}
ds^2&=&g_{tt}dt^2+2g_{ty}dt\,dy+g_{yy}dy^2\\
&&\qquad{}+g_{ij}[dx^i+g^{ik}(A_kdt+B_kdy)][dx^j+g^{j\ell}(A_\ell dt+B_\ell 
dy)].\nonumber
\eea
Just as in the case of the single foliation \cite{mqc,lif}, the appropriate 
action of (\ref{foldbl}) on the fields of (\ref{parmet}) can be obtained 
\cite{sch} by taking a nonrelativistic scaling limit of full spacetime 
diffeomorphisms 
$\diff(M)$, which results from substituting $A_i\to A_i/c$, $B_i\to cB_i$, 
parametrizing the generators of $\diff(M)$ as $(cf,\xi^i,\eta/c)$, and taking 
$c\to\infty$.  This process yields transformation rules for the metric 
components under the action of $\diff(M;\CF_2)$ which are compatible with the 
anisotropic Weyl transformations
\bea
\label{weyldbl}
g_{tt}\to\Omega^{2z}g_{tt},&&\ g_{ty}\to\Omega^{2}g_{ty},\quad 
g_{yy}\to g_{yy},\\
g_{ij}\to\Omega^{2}g_{ij},&&\ A_{i}\to\Omega^{2}A_{i},\quad 
B_{i}\to\Omega^{2}B_{i}. \nonumber
\eea
As we now illustrate in a number of examples, this version of anisotropic Weyl 
transformations again leads to a natural notion of anisotropic conformal 
infinity. 

\subsection{Asymptotic structure of null warped $AdS_3$}

Perhaps the simplest example is null warped $AdS_3$ \cite{nullads}, 
\be
ds^2=-\frac{dt^2}{w^4}+\frac{2dt\,d\theta+dw^2}{w^2}.
\ee
We choose the global scaling of the coordinates to be
\be
t\to\lambda^2t,\qquad w\to\lambda w,\qquad\theta\to\theta.
\ee
This is an example of the scaling defined in (\ref{scaledbl}).  
Using (\ref{weyldbl}) with $\Omega=w$, 
the metric is mapped to
\be
\tilde{ds^2}=-dt^2+2dt\,d\theta+dw^2.
\ee
The anisotropic conformal boundary is again at $w=0$, and satisfies properties 
($a$)-($c$) just like the Lifshitz space, with one novelty:  The group of 
anisotropic conformal symmetries -- defined again as those $\diff(\p M;\CF)$ 
elements that map the boundary metric $-dt^2+2dt\,d\theta$ to itself up to an 
anisotropic Weyl transformation -- is now infinite dimensional, with 
generators
\be
\label{infconf}
F(t)\p_t+G(t)\p_\theta,
\ee
with $F(t)$, $G(t)$ arbitrary.  
Their action on the conformal class of metrics in $\p M$ is induced by 
{\it asymptotic} $\diff(M;\CF_2)$ isometries of null warped $AdS_3$.  In the 
quantum theory, (\ref{infconf}) will give rise to a Virasoro algebra together 
with a $U(1)$ current algebra.

\subsection{Asymptotic structure of the Schr\"odinger space}

Our next example, the Schr\"odinger space
\be
\label{schmet}
ds^2=-\frac{dt^2}{w^{2z}}+\frac{2dt\,d\theta+d\bx^2+dw^2}{w^2},
\ee
has been proposed \cite{son,mcg} as a gravity dual of Galilean-invariant 
nonrelativistic CFTs with dynamical exponent $z$.  In order to get a 
well-behaved anisotropic conformal infinity, we use the scalings of 
(\ref{scaledbl}), with $x^i\equiv(w,\bx)$ and with $y\equiv\theta$ 
an ultralocal dimension.  Using (\ref{weyldbl}) together with $\Omega=w$ 
yields
\be
\tilde{ds^2}=-dt^2+2dt\,d\theta+d\bx^2+dw^2,
\ee
with $\p M$ again at $w=0$.  Note an interesting feature:  Because $\theta$ 
scales with conformal exponent $\zeta=0$, this dimension is present {\it both 
in the bulk and in the boundary}, even if it is compactified; the conformal 
infinity 
% of Schr\"odinger space 
is of codimension one.  This interpretation of 
$\theta$ resolves some of the mysteries associated with this extra bulk 
dimension in % the proposed 
holography of Schr\"odinger spaces.  

The bulk isometries of (\ref{schmet}) again induce the action of anisotropic 
conformal symmetries on the anisotropic conformal class 
$-dt^2+2dt\,d\theta+d\bx^2$ of boundary metrics.  For example, in the case of 
$z=2$, this group of conformal transformations of $\p M$ induced from the 
bulk isometries is generated by
\bea
t^2\p_t+tx^i\p_i-\frac{1}{2}x^2\p_\theta,\qquad&&t\p_i+x^i\p_\theta,\cr
 t\p_t+\frac{1}{2}x^i\p_i,\qquad\p_i,\qquad \p_\theta,&&\qquad x^i\p_j-x^j\p_i.
\eea
These are the generators of the Schr\"odinger conformal group.  
Asymptotic bulk isometries formally extend this symmetry to an 
infinite-dimensional one \cite{ali,compdets,sch}, analogous to (\ref{infconf}).

The metric (\ref{schmet}) describes the Schr\"odinger space in Poincar\'e-like 
coordinates.  At least when $z=2$, it can be analytically continued beyond the 
Poincar\'e patch, to global Schr\"odinger space \cite{blau,sch}
\be
\label{schgl}
ds^2=-\left(1+\frac{\hat\bx^2}{\hat w^2}+\frac{1}{\hat w^4}\right)d\hat t^{\,2}
+\frac{2d\hat td\hat\theta+d\hat\bx^2+d\hat w^2}{\hat w^2}.
\ee
(\ref{schmet}) and (\ref{schgl}) are related by coordinate transformation
\bea
\hat t=\arctan t,\ &&\ \hat\theta=\theta+\frac{t}{2(1+t^2)}(\bx^2+w^2)\cr
\hat x^i=\frac{x^i}{\sqrt{1+t^2}},&&\ \hat w=\frac{w}{\sqrt{1+t^2}}.
\eea
It is reassuring that this transformation is a  double-foliation preserving 
diffeomorphism, of the form (\ref{foldbl}).  As a result, the anisotropic 
conformal boundary of global Schr\"odinger space can also be analyzed in our 
framework.  

\section{Holographic Renormalization and Anisotropic Conformal Infinity}

In the few examples presented above, we simply determined the correct 
form of foliation-preserving diffeomorphisms and the correct 
anisotropic Weyl transformations by inspection.  More complicated examples may 
involve multiple foliations and multiple anisotropies which obscure the 
precise details of the construction.  It is therefore desirable to have 
an algorithmic tool for deriving the anisotropic asymptotic structure in 
more general cases.

We now outline how such rules can be systematically derived from 
considerations of holography in spacetimes with asymptotically anisotropic 
scaling.

\subsection{The general prescription}

The general prescription consists of the following steps:

(i) Identify consistent fall-off conditions on fields on $M$. (This step, 
which we take as an input, is a consequence of the precise definition of the 
dynamics, designed to identify a consistent phase space of the theory; see, 
\eg , \cite{brownh,warpedbh,kerrcft,compdet}. for examples).

(ii) Identify the maximal subgroup of diffeomorphism symmetries compatible 
with (i).

(iii) Identify the anisotropic Weyl transformations compatible with (ii).

Given (i)-(iii), one can then relax the asymptotic fall-off conditions on the 
background to allow for a generic boundary metric $\gamma$, and {\it derive\/} 
the action of the asymptotic symmetries from (ii) on $\gamma$.  This action 
yields the appropriately scaled version of appropriate foliation-preserving 
diffeomorphisms of $\p M$ on the boundary metric $\gamma$.  The anisotropic 
Weyl transformations on $\gamma$ are determined simply from the 
reparametrizations of the radial coordinate.  Finally, we use (iii) to 
construct the anisotropic conformal infinity of $M$.

This general prescription can be illustrated with spacelike warped $AdS_3$ as 
an example.  

\subsection{Asymptotic structure of spacelike warped $AdS_3$}

The metric of the spacelike warped $AdS_3$ in global coordinates is 
\cite{warpedbh}
\be
\label{swarped}
ds^2=-(1+r^2)du^2+\frac{dr^2}{1+r^2}+\frac{4\nu^2}{\nu^2+3}(r\,du+dv)^2.
\ee
Following steps (i)-(iii) outlined above, we get:

(i) Fall-off conditions on the deviations $h_{\mu\nu}$ of the metric from the 
background (\ref{swarped}) were proposed in \cite{compdet}, 
\bea
h_{uu}&=&\CO(r),\qquad h_{vv}=\CO\left(\frac{1}{r}\right),\qquad 
h_{rr}=\CO\left(\frac{1}{r^3}\right),\cr 
h_{uv}&=&\CO(1),\qquad h_{ru}=\CO\left(\frac{1}{r}\right),\qquad 
h_{rv}=\CO\left(\frac{1}{r^2}\right).
\eea

(ii) The group of diffeomorphisms preserving these fall-off conditions is
generated by
\be
\label{asym}
\left[F(u)+\CO\left(\frac{1}{r^2}\right)\right]\p_u-\left[rF'(u)+\CO\left(
\frac{1}{r}\right)\right]\p_r
+\left[G(u)+\CO\left(\frac{1}{r}\right)\right]\p_v,
\ee
with $F(u)$, $G(u)$ arbitrary, and exhibits a natural asymptotic foliation 
structure.  

(iii) Given (\ref{asym}), we choose the anisotropic Weyl transformations 
\be
\label{awiii}
g_{uu}\to\Omega^4 g_{uu},\quad g_{uv}\to\Omega^2 g_{uv},\quad g_{vv}\to g_{vv}
\ee
on the metric.  These are of the form (\ref{weyldbl}), with $z=2$.  Together 
with the asymptotic diffeomorphisms (\ref{asym}), the Weyl transformations 
form an algebra that closes up to subleading terms in $1/r$.  With the choice 
of $\Omega=r^{-1/2}$, we obtain the anisotropic conformal infinity of 
spacelike warped $AdS_3$.  The boundary at anisotropic infinity is 
two-dimensional, and carries an induced anisotropic conformal class of metrics 
represented by 
\be
-du^2+\frac{4\nu^2}{\nu^2+3}(du+dv)^2.
\ee
The action of the correctly scaled form of $\diff(\p M;\CF)$ on the boundary 
metric $\gamma(u,v)$ can be determined by relaxing the background to
\bea
g_{uu}=r^2\gamma_{uu}(u,v)+\CO(r),&&\qquad g_{vv}=\gamma_{vv}(u,v)+\CO\left(
\frac{1}{r}\right),\cr 
g_{uv}=r\gamma_{uv}(u,v)+\CO(1),&&\qquad g_{ru}=\CO\left(\frac{1}{r}\right),
\cr 
g_{rr}=\frac{1}{r^2}+\CO\left(\frac{1}{r^3}\right),&&\qquad 
g_{rv}=\CO\left(\frac{1}{r^2}\right),
\eea
and acting with the group of bulk diffeomorphisms which preserve this 
asymptotic form of the metric, 
\be
\left[F(u)+\CO\left(\frac{1}{r^2}\right)\right]\p_u+ \left[rH(u)+\CO\left(
\frac{1}{r}\right)\right]\p_r
+\left[G(u,v)+\CO\left(\frac{1}{r}\right)\right]\p_v.
\ee
The radial bulk diffeomorphisms generated by $rH(u)\p_r$ induce the correct 
anisotropic Weyl transformation (\ref{awiii}) on the boundary metric, with 
$H$ as the generator.  And the diffeomorphisms along $u$ and $v$ 
precisely reproduce the action of $\diff(\p M;\CF)$ on $\gamma$ that we 
obtained from the $c\to\infty$ scaling %  limit 
below Eqn.~(\ref{parmet})!  
We can use this action of $\diff(\p M;\CF)$ to define the group of 
anisotropic conformal transformations of the boundary metric. This group 
is found to be generated by 
\be
F(u)\p_u+G(u)\p_v.  
\ee
For compact $u$, this reproduces the Virasoro algebra and the $U(1)$ current 
algebra found in \cite{compdet}.

A closely related example is the near-horizon extreme Kerr geometry 
\cite{barhor,kerrcft}, which can be viewed as a family 
of spacelike warped $AdS_3$'s, fibered over the polar coordinate $\theta$.
In this example, $\theta$ is an ultralocal dimension, analogous to 
$\theta$ of the Schr\"odinger space (\ref{schmet}),  but without translational 
invariance along $\theta$.

\section{Conclusions}

The notion of anisotropic conformal infinity clarifies the asymptotic 
structure of vacuum spacetimes with asymptotically anisotropic scaling.  As 
an application, we can now give a precise definition of black holes in 
spacetimes with anisotropic asymptopia:  First, we define an event horizon in 
an asymptotically anisotropic spacetime as the boundary of the causal past of 
the anisotropic infinity, and define black holes as solutions with event 
horizons.  Our definition of anisotropic conformal infinity naturally extends 
to the black holes themselves:  For example, one can show that the spacelike 
warped $AdS_3$ black holes of \cite{warpedbh} share the asymptotic structure 
of the spacelike warped $AdS_3$ vacuum determined above.

In relativistic gravity, the structure of conformal infinity is probed by null 
geodesics.  Spacetimes with anisotropic scaling appearing in the context of 
\cite{mqc,lif} can be similarly probed, by Lifshitz particles \cite{sch} with 
a gapless nonrelativistic dispersion relation. 

Results of this paper illustrate that under pressure from the interface of 
quantum gravity with condensed matter, some of the central notions of 
general relativity must be revisited and adapted for the era in which quantum 
gravity is applied to systems with anisotropic scaling.  In the process, is 
appears that we must disentangle two concepts which seemed so inseparable in 
the physics of the 20th century:  gravity and relativity. 

\acknowledgments 
We wish to thank St\'ephane Detournay for useful discussions.  
The results presented in this paper were announced by one of us (PH) at 
{\it Strings 2009\/} in Rome (June 2009), and at the {\it Quantum Criticality 
and AdS/CFT Correspondence} Miniprogram at KITP, Santa Barbara (July 2009); 
PH wishes to thank the orgainzers for their hospitality.  
This work has been supported by NSF Grants  PHY-0555662 and PHY-0855653, 
DOE Grant DE-AC02-05CH11231, and by the Berkeley Center for Theoretical 
Physics.  

%%%%%%%%%%%%%%%%%%%%%%%%%%%%%%%%%%%%%%%%%%%%%%%%%%%%%%%%%%%%%%%%%%%%%%%%%%%%%%%
\bibliographystyle{JHEP}
\bibliography{aci}

\providecommand{\href}[2]{#2}\begingroup\raggedright\begin{thebibliography}{10}

\bibitem{son}
D.~T. Son, {\it {Toward an AdS/Cold Atoms Correspondence: A Geometric
  Realization of the Schr\"{o}dinger Symmetry}},  {\em Phys. Rev.} {\bf D78}
  (2008) 046003, [\href{http://xxx.lanl.gov/abs/arXiv:0804.3972}{{\tt
  arXiv:0804.3972}}].

\bibitem{mcg}
K.~Balasubramanian and J.~McGreevy, {\it {Gravity Duals for Non-Relativistic
  CFTs}},  {\em Phys. Rev. Lett.} {\bf 101} (2008) 061601,
  [\href{http://xxx.lanl.gov/abs/arXiv:0804.4053}{{\tt arXiv:0804.4053}}].

\bibitem{mqc}
P.~Ho\v{r}ava, {\it {Membranes at Quantum Criticality}},  {\em JHEP} {\bf 03}
  (2009) 020, [\href{http://xxx.lanl.gov/abs/arXiv:0812.4287}{{\tt
  arXiv:0812.4287}}].

\bibitem{lif}
P.~Ho\v{r}ava, {\it {Quantum Gravity at a Lifshitz Point}},  {\em Phys. Rev.}
  {\bf D79} (2009) 084008, [\href{http://xxx.lanl.gov/abs/arXiv:0901.3775}{{\tt
  arXiv:0901.3775}}].

\bibitem{spdim}
P.~Ho\v{r}ava, {\it {Spectral Dimension of the Universe in Quantum Gravity at a
  Lifshitz Point}},  {\em Phys. Rev. Lett.} {\bf 102} (2009) 161301,
  [\href{http://xxx.lanl.gov/abs/arXiv:0902.3657}{{\tt arXiv:0902.3657}}].

\bibitem{holo}
K.~Skenderis, {\it {Lecture Notes on Holographic Renormalization}},  {\em
  Class. Quant. Grav.} {\bf 19} (2002) 5849--5876,
  [\href{http://xxx.lanl.gov/abs/hep-th/0209067}{{\tt hep-th/0209067}}].

\bibitem{gkp}
R.~P. Geroch, E.~H. Kronheimer, and R.~Penrose, {\it {Ideal Points in
  Space-Time}},  {\em Proc. Roy. Soc. Lond.} {\bf A327} (1972) 545--567.

\bibitem{penrin}
R.~Penrose and W.~Rindler, {\em {Spinors and Space-Time. Vol. 2}}.
\newblock Cambridge Univ. Press, 1986.

\bibitem{sch}
P.~Ho\v{r}ava and C.~M. Melby-Thompson, {\it {Anisotropic Conformal Infinity
  and the Global Structure of Schr\"{o}dinger Spaces}},  to appear.

\bibitem{klm}
S.~Kachru, X.~Liu, and M.~Mulligan, {\it {Gravity Duals of Lifshitz-like Fixed
  Points}},  {\em Phys. Rev.} {\bf D78} (2008) 106005,
  [\href{http://xxx.lanl.gov/abs/arXiv:0808.1725}{{\tt arXiv:0808.1725}}].

\bibitem{nullads}
S.~Detournay, D.~Orlando, P.~M. Petropoulos, and P.~Spindel, {\it
  {Three-Dimensional Black Holes from Deformed Anti-de Sitter}},  {\em JHEP}
  {\bf 07} (2005) 072, [\href{http://xxx.lanl.gov/abs/hep-th/0504231}{{\tt
  hep-th/0504231}}].

\bibitem{ali}
M.~Alishahiha, R.~Fareghbal, A.~E. Mosaffa, and S.~Rouhani, {\it {Asymptotic
  Symmetry of Geometries with Schr\"{o}dinger Isometry}},
  \href{http://xxx.lanl.gov/abs/arXiv:0902.3916}{{\tt arXiv:0902.3916}}.

\bibitem{compdets}
G.~Comp\`{e}re, S.~de~Buyl, S.~Detournay, and K.~Yoshida, {\it {Asymptotic
  Symmetries of Schr\"odinger Spacetimes}},
  \href{http://xxx.lanl.gov/abs/arXiv:0908.1402}{{\tt arXiv:0908.1402}}.

\bibitem{blau}
M.~Blau, J.~Hartong, and B.~Rollier, {\it {Geometry of Schr\"{o}dinger
  Space-Times, Global Coordinates, and Harmonic Trapping}},  {\em JHEP} {\bf
  07} (2009) 027, [\href{http://xxx.lanl.gov/abs/arXiv:0904.3304}{{\tt
  arXiv:0904.3304}}].

\bibitem{brownh}
J.~D. Brown and M.~Henneaux, {\it {Central Charges in the Canonical Realization
  of Asymptotic Symmetries: An Example from Three-Dimensional Gravity}},  {\em
  Commun. Math. Phys.} {\bf 104} (1986) 207--226.

\bibitem{warpedbh}
D.~Anninos, W.~Li, M.~Padi, W.~Song, and A.~Strominger, {\it {Warped $AdS_3$
  Black Holes}},  {\em JHEP} {\bf 03} (2009) 130,
  [\href{http://xxx.lanl.gov/abs/arXiv:0807.3040}{{\tt arXiv:0807.3040}}].

\bibitem{kerrcft}
M.~Guica, T.~Hartman, W.~Song, and A.~Strominger, {\it {The Kerr/CFT
  Correspondence}},  \href{http://xxx.lanl.gov/abs/arXiv:0809.4266}{{\tt
  arXiv:0809.4266}}.

\bibitem{compdet}
G.~Comp\`{e}re and S.~Detournay, {\it {Boundary Conditions for Spacelike and
  Timelike Warped $AdS_3$ Spaces in Topologically Massive Gravity}},
  \href{http://xxx.lanl.gov/abs/arXiv:0906.1243}{{\tt arXiv:0906.1243}}.

\bibitem{barhor}
J.~M. Bardeen and G.~T. Horowitz, {\it {The Extreme Kerr Throat Geometry: A
  Vacuum Analog of $AdS_2\times S^2$}},  {\em Phys. Rev.} {\bf D60} (1999)
  104030, [\href{http://xxx.lanl.gov/abs/hep-th/9905099}{{\tt
  hep-th/9905099}}].

\end{thebibliography}\endgroup
\end{document}